# Functional Analysis of Variance for Association Studies

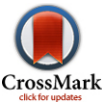

Olga A. Vsevolozhskaya[1]\*, Dmitri V. Zaykin[2], Mark C. Greenwood[3], Changshuai Wei[1], Qing Lu[1]\*

1 Department of Epidemiology and Biostatistics, Michigan State University, East Lansing, Michigan, United States of America, 2 National Institute of Environmental Health Sciences, National Institutes of Health, Research Triangle Park, North Carolina, United States of America, 3 Department of Mathematical Sciences, Montana State University, Bozeman, Montana, United States of America

## Abstract

While progress has been made in identifying common genetic variants associated with human diseases, for most of common complex diseases, the identified genetic variants only account for a small proportion of heritability. Challenges remain in finding additional unknown genetic variants predisposing to complex diseases. With the advance in next-generation sequencing technologies, sequencing studies have become commonplace in genetic research. The ongoing exome-sequencing and whole-genome-sequencing studies generate a massive amount of sequencing variants and allow researchers to comprehensively investigate their role in human diseases. The discovery of new disease-associated variants can be enhanced by utilizing powerful and computationally efficient statistical methods. In this paper, we propose a functional analysis of variance (FANOVA) method for testing the association of sequence variants in a genomic region with a qualitative trait. The FANOVA has a number of advantages: (1) it tests for a joint effect of gene variants, including both common and rare; (2) it fully utilizes linkage disequilibrium and genetic position information; and (3) allows for either protective or risk-increasing causal variants. Through simulations, we show that FANOVA outperform two popularly used methods – SKAT and a previously proposed method based on functional linear models (FLM), – especially if a sample size of a study is small and/or sequence variants have low to moderate effects. We conduct an empirical study by applying three methods (FANOVA, SKAT and FLM) to sequencing data from Dallas Heart Study. While SKAT and FLM respectively detected *ANGPTL 4* and *ANGPTL 3* associated with obesity, FANOVA was able to identify both genes associated with obesity.

**Funding:** This work was supported by a National Institute on Drug Abuse T32 research training program grant award (T32DA021129), QL's NIDA Mentored Research Scientist Development Award (K01DA033346), and by DVZ's Intramural Research Program of the National Institutes of Health, National Institute of Environmental Health Sciences. The content is the sole responsibility of the authors and does not necessarily represent the official views of Michigan State University, the National Institute on Drug Abuse, the National Institute of Environmental Health Sciences or the National Institutes of Health. The funders had no role in study design, data collection and analysis, decision to publish, or preparation of the manuscript.

**Competing Interests:** The authors have declared that no competing interests exist.

\* Email: qlu@msu.edu (QL); vsevoloz@msu.edu (OAV)

## Introduction

Advances in genotyping and sequencing technologies have revolutionized genetic studies of human diseases [1]. During the past decade, genome-wide association studies (GWAS) have identified thousands of common genetic variants predisposing to hundreds of human diseases [2]. Nevertheless, for most complex diseases, the genetic variants identified from GWAS only explain a small proportion of heritability [3]. Additionally, a robust replication of initial genetic association findings has proved to be difficult. There is a number of explanations for a lack of reproducible results: population stratification, variability in linkage disequilibrium among samples, and, more importantly, weak genetic effects and lack of statistical power. The traditional single-marker association studies can detect variants that explain a relatively large fraction of the phenotypic difference, however, the effect size of newly discovered variants is typically overestimated [4]. A single strong association suggested by initial studies often has a more subtle disease protection or predisposition effect in the subsequent research. Moreover, the majority of genetic effects are expected to be very small, resulting in an L-shaped distribution of effect sizes [5,6] and studies of small size are under-powered to reliably detect them. Many variants in a study might be truly associated, but not detectable or not reproducible by a single-marker analysis.

To identify additional genetic variants contributing to human diseases, efforts have been shifted from common variants of small effect towards uncovering rare variants of large effect. Benefiting from the recent high-throughput exome sequencing and whole-genome sequencing technologies, researchers were able to assess the spectrum of rare variants and comprehensively study their role in human diseases [7,8]. Sequencing studies have been successful in identifying genetic variants contributing to Mendelian disorders and undiagnosed childhood genetic diseases [9]. Nevertheless, challenges remain in uncovering genetic variants associated with complex human diseases.

To improve the power of association analysis with sequencing data, burden tests, such as the weighted sum test [10], have been introduced to evaluate the combined effect of multiple variants in a genetic region. By collapsing information over multiple variants, burden tests attain improved performance for sequencing data analysis and alleviate the multiple-testing burden. However, the limitation of a burden test lies in its assumption of the same direction of the effects [11]. Such an assumption may not hold in certain scenarios (e.g., a gene with both loss-of-function and gain-of-function mutations). To relax this assumption, many methods have been developed, including the C-alpha test [12], the sequence kernel association test (SKAT) [11], and the estimated regression coefficient test [13].

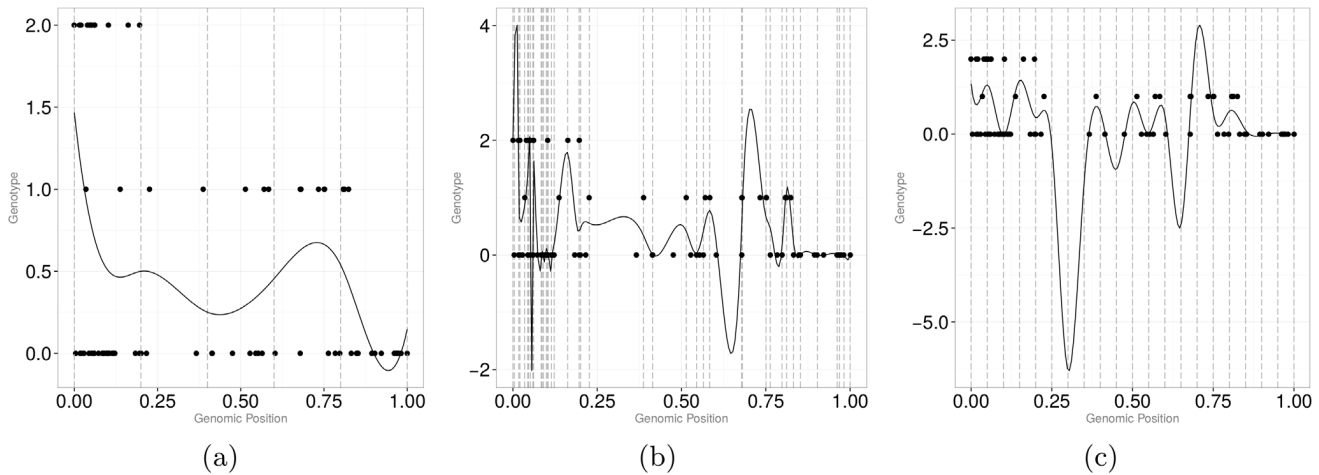

**Figure 1. Smooth curves obtained by cubic B-splines using six knots (*left panel*), 42 knots placed at every other variant position (*middle panel*), and 21 equally spaced knots (*right panel*).**
doi:10.1371/journal.pone.0105074.g001

It is now increasingly recognized that many variants jointly contribute to the observed phenotypic variation in complex human traits [6,14] and it would be desirable to study their joint effect without having to specify an arbitrary threshold to collapse information over rare variants or a weight function to combine common and rare variants. To address this issue, we propose the use of the functional analysis of variance (FANOVA) for association analysis using sequence data by testing for an overall difference in allele frequencies of multiple variants within a genetic locus between unaffected and affected populations. A locus is defined as a region of continuous sequence that includes many (both common and rare) variants. Similar to other function-valued methods, such as a test based on functional linear models (FLM) [15], the word "functional" in FANOVA indicates that the method is based on the analysis of continuous functions (curves) – an approach to analyzing high-dimensional data popularized by, among others, Ramsay and Silverman [16]. As a function-valued method, FANOVA has a number of attractive features, including the built-in facility to account for the correlation structure (i.e., linkage disequilibrium, LD), absence of restrictive assumptions, such as that rare variants confer a deleterious effect direction as well as a relatively higher risk compared to common variants, and no requirement of a weight function or a threshold to handle rare variants. Additionally, the functional *analysis of variance* fits perfectly with the paradigm of most case-control association studies, and has the advantage over an existing function-valued method [15] when the sample size of a sequencing study is small and/or variants have low to moderate effects.

For both the currently proposed FANOVA approach and the FLM method of Luo et al. [15], genetic data (i.e., genotype labeling) of an individual need to be represented by a smooth function. Therefore, we begin with a discussion of how to build functional data objects and propose a genotype relabeling method that provides data that more coincide with smooth characteristics required to treat observations as functions. We then introduce FANOVA and provide a brief overview of SKAT and FLM. A simulation study is conducted to evaluate empirical type I error rates and power of these methods. We conclude with real data application to the Dallas Heart sequencing data and a subsequent discussion.

## Materials and Methods

### Smoothing

Before applying functional methods to perform an association test, we need to transform the discrete genotype profile of a subject (i.e., sequencing variants typically coded as 0, 1 or 2) into smooth functions $y(t)$, measured at $t = 1, \ldots, T$ allele genomic positions. Smooth functional representation is achieved by the use of basis functions and is discussed in several textbooks, e.g., [16–18]. Here, we provide a brief overview of how to reconstruct smooth functional data from discrete observations and caution a reader of potential pitfalls encountered in the application of smoothing methods to genomic data.

In functional data analysis, we denote a function, $y(t)$, as a linear combination of independent basis functions, $\phi_k(t)$, with coefficients, $b_k$, as

$$y(t) = \sum_{k=1}^{K} b_k \phi_k(t). \tag{1}$$

The choice of basis functions is rather wide but can be narrowed to two popular options: Fourier basis system (i.e., $1$, $\sin(\omega t)$, $\cos(\omega t)$, $\sin(2\omega t)$, $\cos(2\omega t)$, $\ldots$, $\sin(K\omega t)$, $\cos(K\omega t)$) for periodic data and a spline basis system using locally defined polynomials for non-periodic data. Because it is difficult to assume that genetic variants exhibit periodic behavior, a spline basis system becomes a natural choice. To specify a spline basis system, one needs to 1) divide the genetic region into sub-intervals by specifying break points; 2) specify the order of the polynomials that are going to be fit on each sub-interval, and 3) define a sequence of knots that are placed at sub-interval breakpoints. We can use Figure (1) to conceptualize the idea. Dots in each panel of Figure (1) represent the same genetic data – 81 variants coded based on the number of minor alleles – versus genomic position of the variants. The genomic region was scaled to [0,1] (the x-axis), which does not influence graphical representation of the data. Vertical dotted lines show "knots" – break points that divide the range of variant positions into segments along with two end points. Cubic B-splines [19] are fit within each segment under the constraint that the fitted line, including the first and second derivatives, is continuous at the break points.

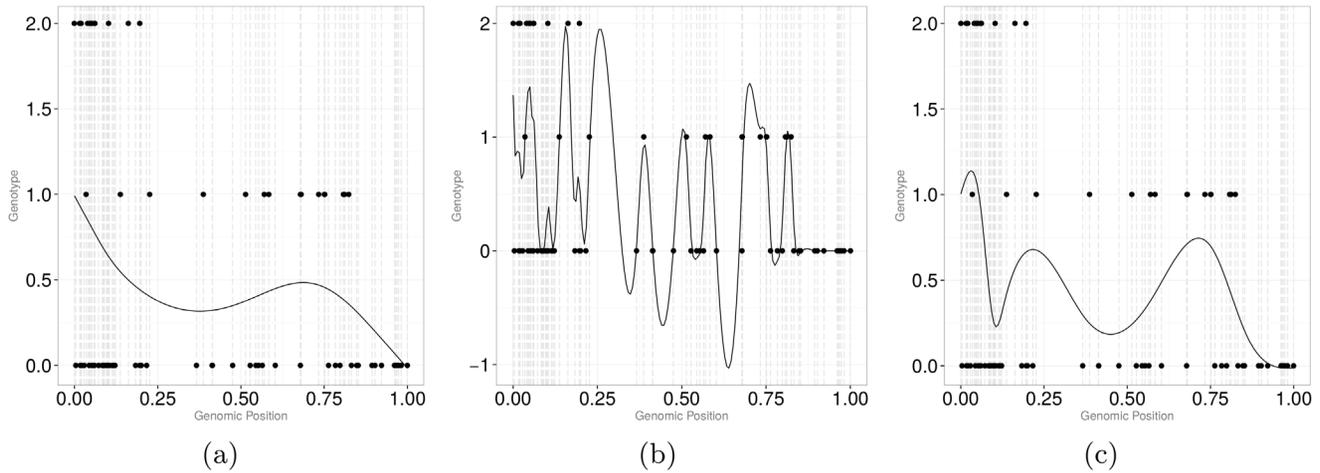

**Figure 2. Smooth curves for different values of $\lambda$.** *Left panel*: $\lambda = 1e-2$, *middle panel*: $\lambda = 1e-7$, and *right panel*: $\lambda$ is determined through generalized cross validation (GCV).
doi:10.1371/journal.pone.0105074.g002

In Figure (1a), the number and the position of knots concur with recommendations in Luo et al. [15]. In Figure (1b), we put a knot in every second position. In Figure (1c), we chose 21 equally spaced knots as one of the values in a range of 10 to 25 recommended in Fan et al. [20]. It is clear that the more knots are used, the less smooth the curve becomes. Additionally, the position of the knots plays a crucial role. When genotypes are missing (as in Figure (1) from 0.25 to about 0.4), then a "larger" value of equally spaced knots may produce a fit to data that has wild excursions over segments with missing information (consider a "wild" dip to over negative five in Figure (1c)).

It is generally unclear how to choose the "optimal" number of knots. A common recommendation is to choose the amount of knots based on the visual comparison of the smoothers. Clearly, this approach is unrealistic with large-scale genomic data, where an "adequate" representation of the discrete data by a smoother is hard to define. Alternatively, the optimal number of knots can be chosen based on AIC (Akaike information criterion) optimization. Wood [17], however, argues that it is unwise to go the route of finding the optimal number of knots. Instead, smoothing splines

(also called penalized splines) can be used. With smoothing splines, a parameter $\lambda$ – a smoothing parameter – can be used to vary the smoothness of a curve for the same number of knots by penalizing the roughness of the curves in penalized least squares estimation. Figure (2) illustrates the effect $\lambda$ on the functional fit. The larger the value of $\lambda$ is, the more linear is the fit. Thus, instead of looking for the optimal number of knots, one may look for the optimal amount of smoothing for a fixed, typically large number of knots. The optimal value for $\lambda$ is usually found using generalized cross validation, GCV [17].

The number of knots and the smoothing parameter aside, the functional fit to the data will depend on the variant coding. Single nucleotide polymorphism (SNP) genotype is usually coded "0" for homozygote with respect to the major allele, "1" for heterozygote, and "2" for homozygote with respect to the minor allele. The power of functional procedures may depend on the linearity of the functional fit. That is, if in the original sequence data the pattern 0-2 (or 2-0) is persistent, which can be expected with a negative pairwise LD between biallelic variants, a smoother is going to have a high frequency of oscillations. Negative LD is expected to be common based on the population genetics theory [21] and is indeed abundant in real data [22]. These oscillations, in turn, can result in a loss of statistical power to detect true signals because of a poor fit of the smoother to the data. To address this issue, one can use a small number of knots as in Luo et al. [15] and mask the LD patterns by a large amount of smoothing. However, we find this solution unsatisfactory and propose an alternative approach. Since the minor allele is not guaranteed to be deleterious, genotype coding should not depend on which of the two alleles is the minor allele. That is, if all "2" values are flipped to "0" values (and vice verse) at a particular genomic position across all cases and controls, the result of association analyses that do not make *a priori* assumptions about the direction and the magnitude of effects should be unaffected. Thus, we suggest a genotype relabeling algorithm to minimize the number of "flips" that, in turn, should decrease the number of noisy oscillations in smoothed data. Figure (3) illustrates the idea by providing sample genotype data for seven subject across two SNPs. Left panel of the table highlights subjects for which relabeling of $SNP_2$ would obliterate the flip; the right panel shows the same data but genotype coding at the $SNP_2$ position relabeled.

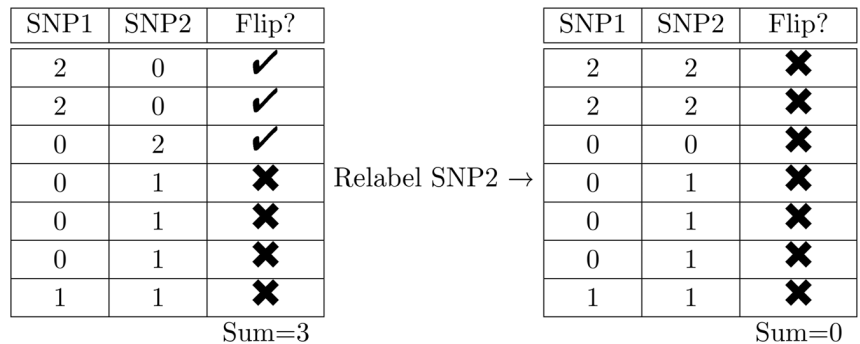

**Figure 3. Sample genotype information over seven subjects and two SNPs.** *Left panel* illustrates original data and highlights subjects for which a flip in genotype coding at $SNP_2$ position is possible. *Right panel* illustrates "flipped" data at the $SNP_2$ position.
doi:10.1371/journal.pone.0105074.g003

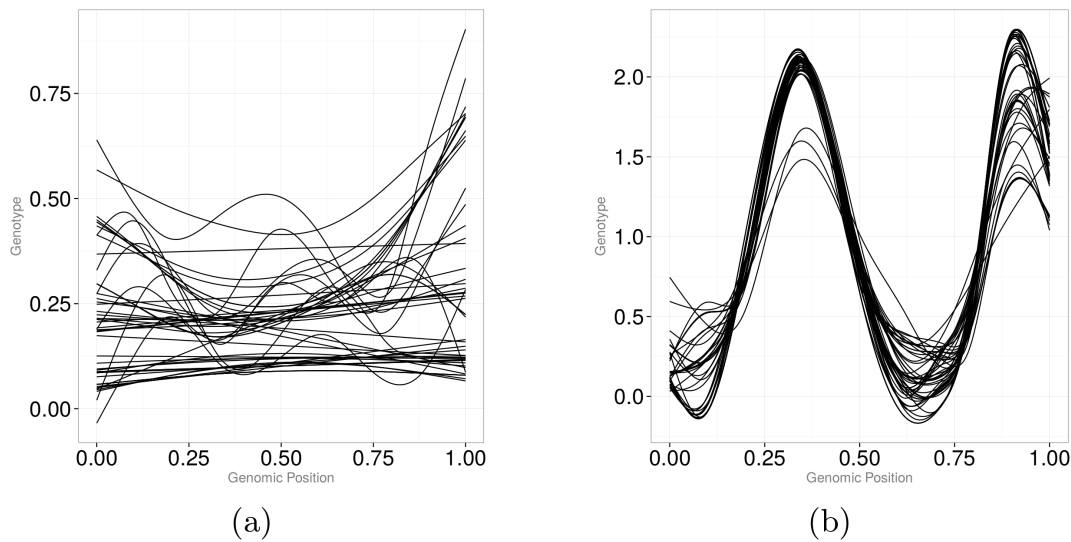

(a)  (b)

**Figure 4. Smooth genotype curves using original data (*left panel*), and relabeled data (*right panel*).**
doi:10.1371/journal.pone.0105074.g004

Figure (4) shows the effect of relabeling on the genotype curves. For the genotype curves, the effect of relabeling is most apparent if GCV is used to determine the optimal amount of smoothing for each subject. Intuitively, this result makes sense. Specifically, with a small number of knots (e.g., Figure (1a)), the functional fit is "too smooth" and is insensitive to arbitrary relabeling. However, whenever the amount of smoothing can vary from subject to subject, and is optimized using the GCV algorithm, the functional fit is sensitive to the relabeling. Similar to continuous registration [16], we expect an increase in statistical power of functional procedures after genotype relabeling (if the functional fit is sensitive to the relabeling) and confirm this result in our simulation study.

## Statistical Procedures

Once the continuous functions are obtained, the question remains how to perform statistical inference with functional data. Here, we propose a functional analysis of variance approach and provide an overview of the methods based on functional linear models by Luo et al. [15], as well as the kernel sequencing association test by Wu et al. [11].

**FANOVA.** Suppose we have $k$ independent groups with functional samples $y_{i1}(t), \ldots, y_{in_i}(t)$, $i = 1 \ldots k$. In functional data

**Table 1.** Estimated type I error rates for each association testing procedure based on 1,000 samples from the null hypothesis with no risk variants at $\alpha = 5\%$ and $N = 500$ subjects.

| Method | Implementation Details | Common Variants | Common and Rare Variants |
|---|---|---|---|
| | Small Basis Cubic Splines | 0.046 | 0.056 |
| | | **0.044** | **0.055** |
| FANOVA | Large Basis Cubic Splines | 0.069 | 0.065 |
| | | **0.065** | **0.063** |
| | Large Basis Penalized Splines | 0.053 | 0.045 |
| | | **0.050** | **0.053** |
| FLM | | 0.042 | 0.049 |
| | | **0.040** | **0.056** |
| SKAT | | 0.045 | 0.044 |

The rows highlighted in bold report type I error rates of the procedures after SNP genotype was relabeled to minimize the number of "2-0" or "0-2" flips.
doi:10.1371/journal.pone.0105074.t001

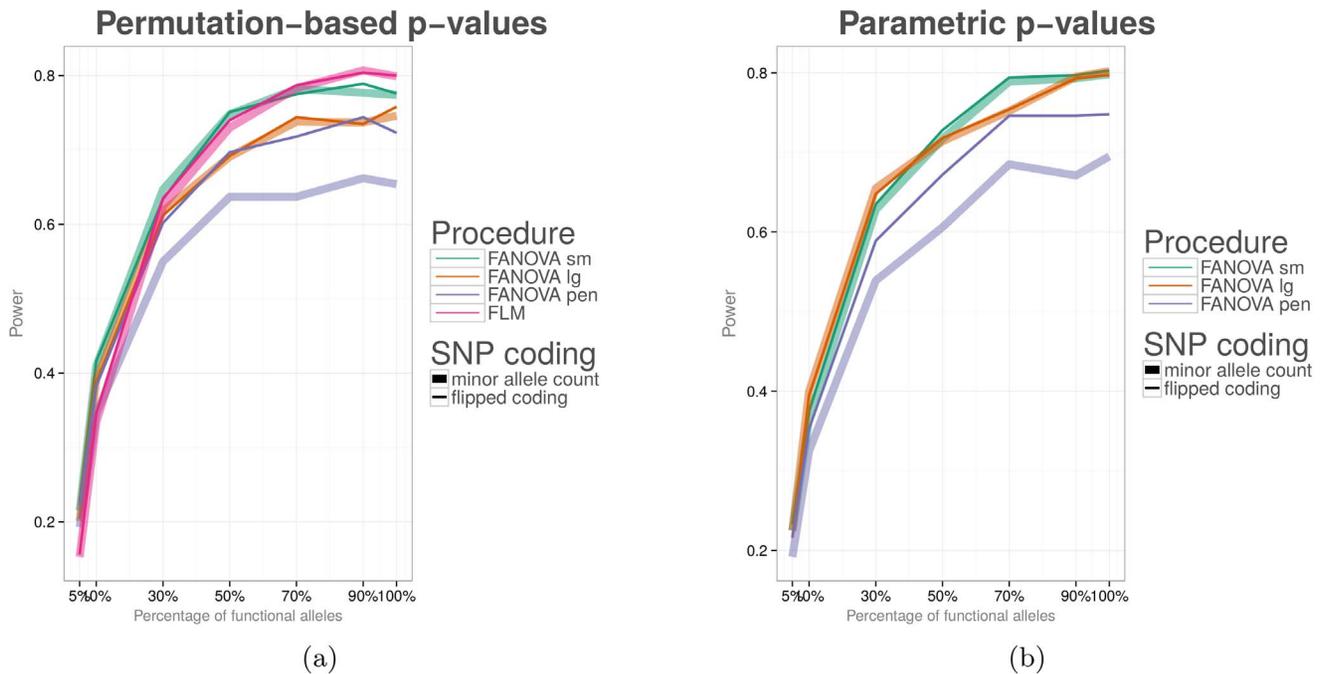

**Figure 5. The effect of the proposed genotype relabeling method on power of functional procedures.** Functional procedures were used to test for an association between all variants (both common and rare) in a genomic region with a dichotomous phenotype. In the **left panel**, statistical power is calculated based on a permutation tests. In the **right panel**, statistical power is based on the parametric test.
doi:10.1371/journal.pone.0105074.g005

analysis, $t$ typically represents a real-valued time variable, but in the current problem it denotes genomic position of a variant; and $y(t)$ is the smoothed genotypic value. We assume that $y_{i1}(t), \ldots, y_{in_i}(t) \overset{i.i.d.}{\sim} \text{GP}(\mu_i(t), \gamma), i = 1, \ldots, k$, where GP stands for "Gaussian Process", $\mu_i(t)$ is unknown mean function and $\gamma(s,t)$ is the common covariance function. The FANOVA model can be written as

$$y_{ij}(t) = \mu_i(t) + \epsilon_{ij}(t), \ \epsilon_{ij}(t) \overset{i.i.d.}{\sim} \text{GP}(0, \gamma)$$

$$j = 1, \ldots, n_i; \ i = 1, \ldots, k.$$

We wish to test if the mean genotype functions vary among $k$ groups over a continuous sequence region $\mathcal{T}$:

$$H_0: \quad \mu_1(t) = \cdots = \mu_k(t), \text{ for all } t \in \mathcal{T},$$

$$H_a: \quad \mu_i(t) \neq \mu_j(t), \text{ for at least one } i \neq j \text{ and } t \in \mathcal{T}.$$

A number of test statistics were proposed to perform the above test (e.g., [23] and [24]). For example, similar to the regular analysis of variance, one can compare between and within group variations:

$$\mathcal{F} = \frac{\int_{\mathcal{T}} \sum_{i=1}^{k} n_i (\hat{\mu}_i(t) - \hat{\mu}(t))^2 dt / (k-1)}{\int_{\mathcal{T}} \sum_{i=1}^{k} \sum_{j=1}^{n_i} (y_{ij}(t) - \hat{\mu}_i(t))^2 dt / (n-k)}, \quad (2)$$

$$= \frac{\int_{\mathcal{T}} \sum_{i=1}^{k} n_i (\hat{\mu}_i(t) - \hat{\mu}(t))^2 dt / (k-1)}{\int_{\mathcal{T}} \hat{\gamma}(t,t) dt},$$

where the group mean functions and the common covariance function are estimated as

$$\hat{\mu}_i(t) = n_i^{-1} \sum_{j=1}^{n_i} y_{ij}(t), \ i = 1, \ldots, k,$$

$$\hat{\mu}(t) = n^{-1} \sum_{i=1}^{k} n_i \hat{\mu}_i(t), \ n = \sum_{i=1}^{k} n_i,$$

**Table 2.** Empirical type I error rates for each association test at the nominal $\alpha = 0.05/50$ based on 10,000 simulations.

| Nominal level | FLM | SKAT | FANOVA |
|---|---|---|---|
| $\alpha = 0.001$ | 0.0007 | 0.0005 | 0.0014 |

doi:10.1371/journal.pone.0105074.t002

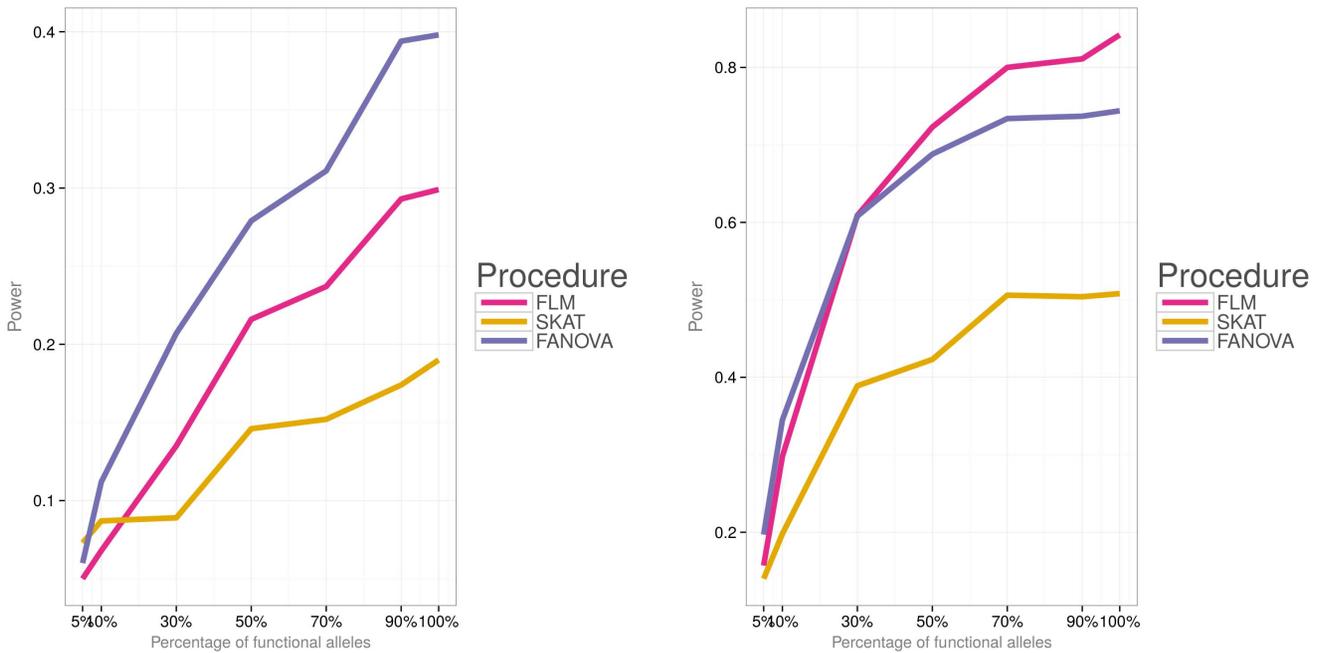

**Figure 6. Empirical power of the three methods for $N=50$ subjects, the first disease model (i.e., $\mu_\beta=0$). Left panel:** small effect $\sigma_\beta=0.25$. **Right panel:** large effect $\sigma_\beta=1$.
doi:10.1371/journal.pone.0105074.g006

$$\hat{\gamma}(s,t) = (n-k)^{-1} \sum_{i=1}^{k} \sum_{j=1}^{n_i} (y_{ij}(s) - \hat{\mu}_i(s))(y_{ij}(t) - \hat{\mu}_i(t)).$$

Under the Gaussian assumptions, it can be shown that the numerator of $\mathcal{F}$ follows a mixture of chi-squared distributions, $\sum_{r=1}^{m} \lambda_m \chi^2_{k-1}$; and the denominator follows a mixture of $\sum_{r=1}^{m} \lambda_m \chi^2_{n-k}$, where $\lambda_1 \geq \lambda_2 \geq \ldots \geq \lambda_m > 0$ are the decreasingly ordered positive eigenvalues of $\gamma(s,t)$ and $\chi^2_{(\cdot)}$ are independent chi-squared random variables. The proof of this result can be found in

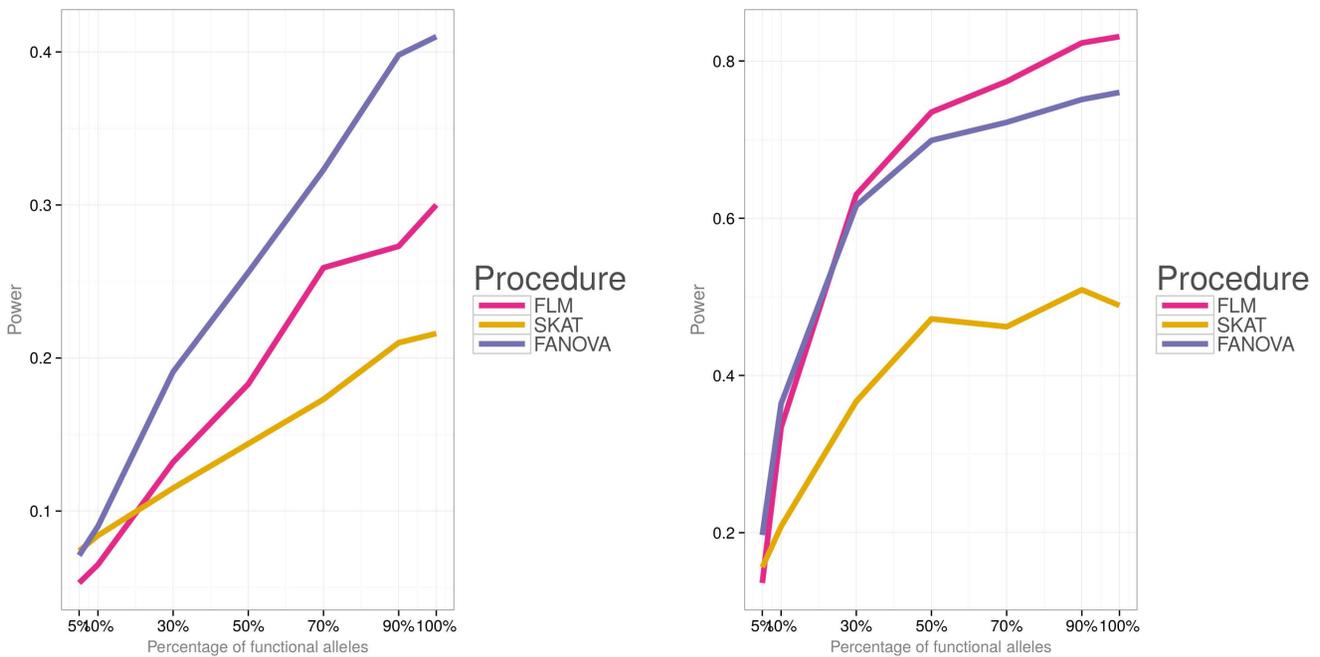

**Figure 7. Empirical power of the three methods for $N=50$ subjects, the second disease model (i.e., $\mu_\beta=0.05$). Left panel:** small effect $\sigma_\beta=0.25$. **Right panel:** large effect $\sigma_\beta=1$.
doi:10.1371/journal.pone.0105074.g007

Zhang [25] Theorem 5.8. By applying the Satterwaite approximation method, one can write

$$\sum_{r=1}^{m} \lambda_m \chi_{(\cdot)}^2 \sim c\chi_{(\cdot)\kappa}^2, \text{ where } c = \frac{\sum_{i=1}^{\infty} \lambda_i^2}{\sum_{i=1}^{\infty} \lambda_i} \text{ and } \kappa = \frac{(\sum_{i=1}^{\infty} \lambda_i)^2}{\sum_{i=1}^{\infty} \lambda_i^2}.$$

The distribution of $\mathcal{F}$ is then approximated as a ratio of

$$\frac{c\chi_{(k-1)\kappa}^2}{c\chi_{(n-k)\kappa}^2} \sim F_{(k-1)\kappa,(n-k)\kappa}. \tag{3}$$

Shen and Faraway [23] denotes $\kappa$ as the "degrees-of-freedom-adjustment-factor". In practice, the continuous functions $y_{ij}(t)$ are discretized on a large grid of $M$ points, $y_{ij}(t_1), \ldots, y_{ij}(t_M)$ and then $\kappa$ can be estimated as $\hat{\kappa} = \mathrm{tr}^2(\hat{\Gamma})/\mathrm{tr}(\hat{\Gamma}^2)$, where $\hat{\Gamma} = \{\hat{\gamma}(t_i, t_j)\}$ is the empirical $M \times M$ covariance matrix. Alternatively, the empirical distribution of $\mathcal{F}$ can be approximated using permutations.

**FLM.** The method proposed by Luo et al. [15] is also based on genotype functions. Unlike the FANOVA approach, which models a function-valued response and a scalar explanatory variable, the FLM approach deals with a scalar response and a function-valued explanatory variable. Specifically, the FLM model is written as:

$$y_i = \beta_0 + \int_{\mathcal{T}} X_i(t)\beta(t)dt + \epsilon_i, \; \epsilon_i \overset{i.i.d.}{\sim} N(0,\sigma), \tag{4}$$

where $y_i$ is the phenotype of subject $i$, $X_i(t)$ is the genotype function, and $\beta(t)$ is the genetic additive effect of a variant at a position $t$. The goal is to test for an association between a phenotypic trait and a genomic region $\mathcal{T}$:

$$H_0: \quad \beta(t) = 0, \text{ for all } t \in \mathcal{T},$$

$$H_a: \quad \beta(t) \neq 0, \text{ for at least one } t \in \mathcal{T}.$$

We now have to estimate the infinite-dimension function $\beta(t)$ based on a finite number of observations $n$, which is an impossible problem. A way around this, is to rewrite $\beta(t)$ in terms of the basis function expansion as in Eq. (1), that is $\beta(t) = \sum_{k=1}^{K_\beta} b_k \theta_k(t)$. This is a dimension reduction technique that simplifies the problem to the standard multiple regression problem. The hypotheses of interest can be build on the coefficients of the basis functions:

$$H_0: \quad b_k = 0 \text{ for all } k = 1, \ldots, K_\beta,$$

$$H_a: \quad b_k \neq 0; \text{ for at least one } k.$$

The coefficient vector **b** can be estimated using the least squares approach, which is detailed in Luo et al. [15] as well as in Ramsay and Silverman [16]. The test statistic is the simple Wald test statistic:

$$T_Q = \hat{\mathbf{b}}^T Var^{-1}(\hat{\mathbf{b}})\hat{\mathbf{b}}, \tag{5}$$

where $\hat{\mathbf{b}}$ is the least squares estimate of the parameter vector **b** and $Var(\hat{\mathbf{b}})$ is the sampling variance of **b**. The Wald test statistic follows a chi-squared distribution, $\chi^2_{K_\beta}$. Nevertheless, because we need to account for selecting a smoothing parameter as well as the smoothing itself, Luo et al. [15] use permutations to obtain the p-values. That is, they first find the observed p-value based on the $\chi^2_{K_\beta}$ distribution, permute the phenotype vector $y_i$ a large number of times and recalculate the p-value based on the $\chi^2_{K_\beta}$ distribution across permutations. The p-value for the test can be obtained as the proportion of the permuted p-values that exceeded the observed p-value.

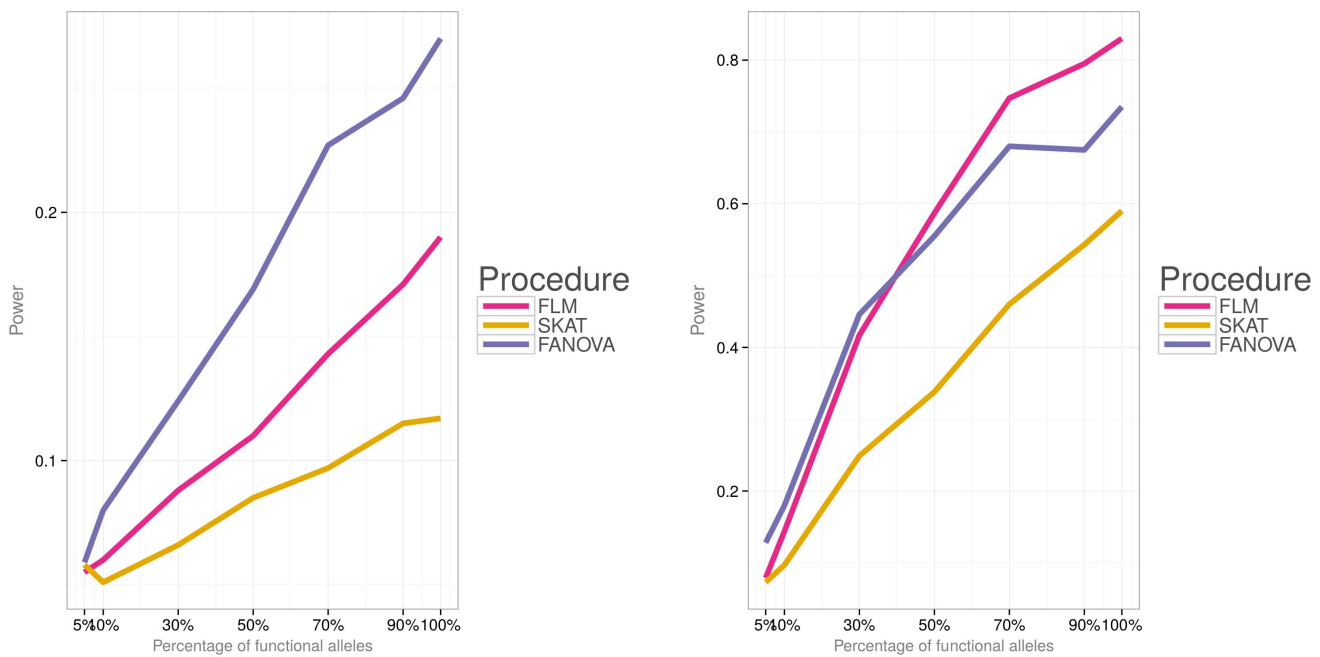

**Figure 8. Empirical power of the three methods for** $N = 500$ **subjects, the first disease model (i.e.,** $\mu_\beta = 0$**). Left panel:** small effect $\sigma_\beta = 0.05$. **Right panel:** large effect $\sigma_\beta = 0.15$.
doi:10.1371/journal.pone.0105074.g008

It should be noted that in Luo et al. [15] the scalar response, $y_i$, was assumed to be a continuous trait and in our application it is a dichotomous variable. However, we can use a linear model in Eq. (4) with a binary phenotype to perform the chi-squared test for an association. In fact, the equivalence of the chi-squared test and the ANOVA test for dichotomous populations was shown by D'Agostino [26] and in our simulations we confirm the adequacy of this approach for the function-valued methods.

**SKAT.** The sequence kernel association test (SKAT) was proposed by Wu et al. [11] as a computationally efficient semi-parametric method to test an association between a SNP set in a region and a continuous or dichotomous trait. The details of SKAT were discussed in multiple publications (e.g., [11,27]) so we leave it to the interested reader to review the mathematical machinery behind it. In short, SKAT draws together single-variant association statistics to compute a p-value for an entire set of SNPs. An interesting result was provided in Kinnamon et al. [22], where they showed an equivalence between the weighted sum of single-variant Cochran-Armitage trend chi-square statistics and the SKAT statistic under an additive genetic model without covariates, and with score statistic weights equal to the inverse estimated null variance of the single-variant Cochran-Armitage trend score statistic.

## Results

We performed simulation studies to evaluate FANOVA and compare its size and power to FLM and SKAT. The genetic data were simulated using the 1,000 genome project [28] to mimic the real sequencing data structure, e.g., realistic linkage disequilibrium pattern, allele frequencies, and randomly missing genotype data. The binary phenotype $X_i$ was simulated using logistic regression as

$$\text{logit}(\Pr(X_i = 1)) = G_i(\text{SNP}_j) \times \beta_j, \qquad (6)$$

where $G_i$ was a genotype label (i.e., 0, 1, or 2) of a causal variant in position $j$; and $\beta_j$, the effect size, was simulated from a normal distribution, $N(\mu_\beta, \sigma_\beta^2)$. The sample size ranged from 50 (25 cases and 25 controls) to 1,000 (500 cases and 500 controls). The simulations were based on variants from a randomly selected 30 kb genetic region. After excluding variants whose genotype labels were constant across all subjects (both cases and controls), each simulated data had an average of 155 variants for 50 subjects, 315 variants for 500 subjects, and 377 variants for 1,000 subjects. For each data set, we tested for an association between a set of all variants (both common and rare) and a dichotomous phenotype.

Two disease models were considered. First, we ran simulations for a model with both deleterious and protective causal variants. Therefore, the effect, $\beta$, was sampled from $N(0, \sigma_\beta^2)$. Second, we considered a model in which the majority of causal variants was assumed to be deleterious and $\beta$ was simulated from $N(\mu_\beta, \sigma_\beta^2)$, with $\mu_\beta > 0$. The strength of the effect size was varied by manipulating $\sigma_\beta^2$. The percentage of causal variants ranged from 5% to 100%.

## Details of the Analysis

To explore the effect of smoothing on functional procedures, we considered three different smoothing strategies:

1. The number and the position of knots were chosen to concur with recommendations in Luo et al. [15] and cubic B-splines were fit within each segment. As illustrated by Figure (1a), this functional fit corresponds to high amount of smoothing.

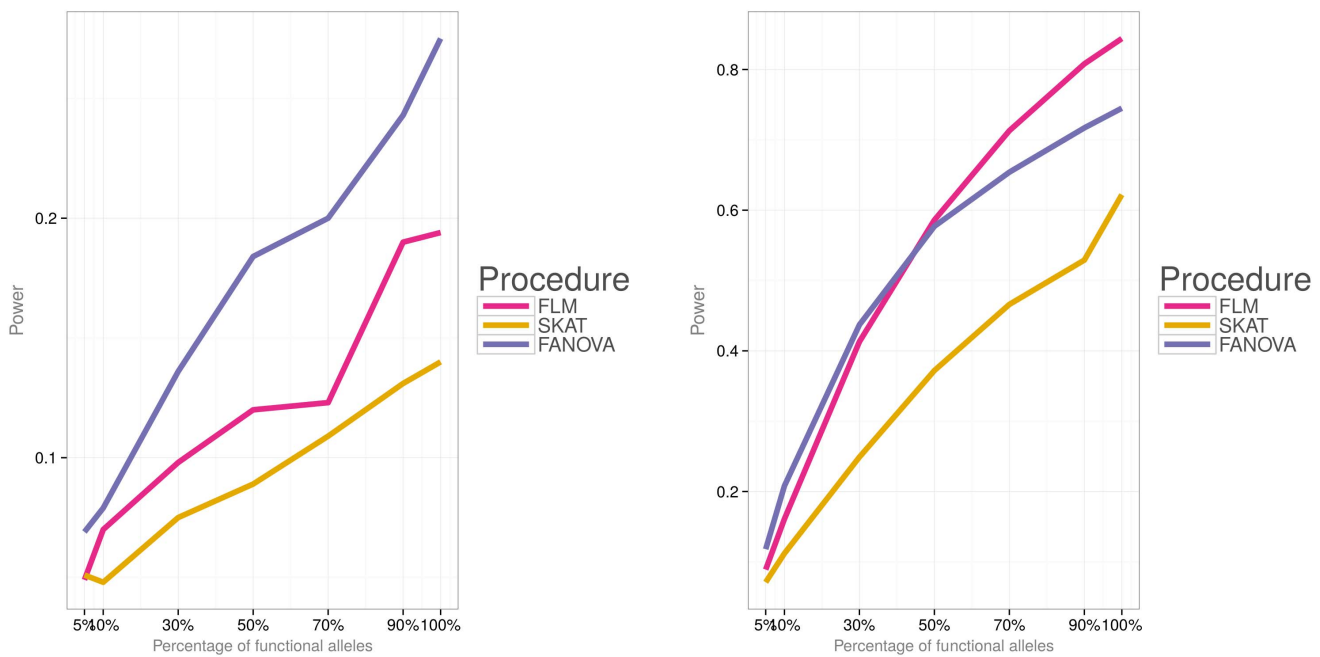

**Figure 9. Empirical power of the three methods for $N = 500$ subjects, the second disease model (i.e., $\mu_\beta = 0.01$). Left panel:** small effect $\sigma_\beta = 0.05$. **Right panel:** large effect $\sigma_\beta = 0.15$.
doi:10.1371/journal.pone.0105074.g009

**Table 3.** P-values of the three methods for testing an association with BMI.

| Gene | FLM | SKAT | FANOVA |
|------|-----|------|--------|
| ANGPTL 3 | 0.046 | 0.075 | 0.037 |
| ANGPTL 4 | 0.069 | 0.003 | 0.002 |
| ANGPTL 5 | 0.864 | 0.721 | 0.441 |
| ANGPTL 6 | 0.927 | 1.000 | 0.754 |

doi:10.1371/journal.pone.0105074.t003

2. A knot was placed in every other genetic position and cubic B-splines were used as basis functions. As illustrated by Figure (1b), the fit results in oscillating curves.

3. A knot was placed in every genetic position and an optimal smoothing parameter, $\lambda$, was chosen separately for each subject through the GCV algorithm. Penalized cubic B-splines are used as basis functions (Figure 1c).

The functional data were estimated using the fda [29] R package [30] with the smoothing parameter, $\lambda$, for the third scenario estimated separately for each subject via mgcv package [31]. The smoothing parameter, determined by the GCV algorithm, ranged from $1 \times 10^{-15}$ to 45.83. The upper bound of $\lambda$ values might seem to be high, but it just indicates that sometimes all simulated genotype labels of a subject were the same, resulting in a linear fit and a high value of lambda. To study the effect of genotype relabeling, we performed an association test with genotype labeling based on the number of minor alleles and for the "flipped" labeling. The association test proposed by Luo et al. [15] based on functional linear models was implemented using the script provided at https://sph.uth.edu/hgc/faculty/xiong/software-F.html. The sequence kernel association test proposed by Wu et al. [11] was implemented using R SKAT package [32] with a small sample adjustment.

## Simulation Study Results

**Empirical Type I Error Rates.** Under the null model with no casual variants, empirical type I error rates are reported in Table 1 at the $\alpha = 5\%$ nominal level. The results were calculated based on 1,000 simulated data sets. In tables, the rows highlighted in bold correspond to the type I errors of the procedures after genotype relabeling that minimizes the number of "flips" was implemented. For $N = 50$ subjects, we calculated p-values based on both the permutation and asymptotic methods (Eq.(2)-(3)). The resolution of the functional discretization, $M$, was twice the number of variants under consideration. Discretized points, $t_1, \ldots, t_M$, were equally spaced over the locus range $\mathcal{T}$. Permutation-based p-values were calculated by shuffling the distance matrix $I = 999$ times which is a computationally more efficient equivalent of the usual permutation procedure where the affection status labels are shuffled. Specifically, the FANOVA test statistic can be expressed in terms of a distance matrix, i.e., symmetric matrix of pairwise dissimilarities among every pair of subjects. In terms of the test statistic values, a simultaneous permutation of rows and columns of the distance matrix is equivalent to a permutation of rows of the data set where each row holds genotype values of an individual. That is, row labels of the distance matrix become exchangeable units allowing one to avoid recalculation of pairwise distances for each permutation of genotype profiles, which greatly reduces computation time [33, 34]. P-values were calculated as the proportion of $I = 999$ recalculated test statistics greater than the observed one [35]. Despite the gains in computational performance due to use of the distance matrix, permutation testing can still be time-consuming as the number of subjects increases. Running time for obtaining one permutation-based p-value for 500 subjects on 2.5 GHz second generation Intel quad-core processor with 8 GB system memory was 68 seconds versus 1.4 second for the asymptotic p-value. As demonstrated by the simulations (i.e., Table 1 and Figure 5), the permutation and asymptotic approximations have comparable performance even when sample sizes are small (i.e., $N = 50$). We therefore only reported asymptotic p-values for FANOVA.

Table 1 shows that for all methods empirical type I error rates are around the nominal 5% level, but for the situations whenever cubic B-splines with a knot in every other position was used to fit smooth curves and p-values were found using parametric approximation. This inflation in type I errors is an artifact of the smoothing method. Genotype relabeling has little effect on the type I error rates. Under the null hypothesis, this result is expected since the size of a test should not depend on how we code a reference allele. The results for SKAT after genotype relabeling are not presented. By default, if SNPs are not coded as minor allele counts, SKAT flips them back to minor allele coding.

Finally, via anonymous review of this article, we were encouraged to check the FANOVA performance at a nominal $\alpha$ level less than 0.05, since in practice multiple genes/regions will be tested. To keep the number of additional simulations feasible, we chose to use the alpha level $0.05/50 = 0.001$, which would correspond to the Bonferroni adjusted alpha for 50 independent regions. We only checked the size of FANOVA for the asymptotic test, since it is the one primarily utilized in this paper. The number of subjects was kept low, $N = 50$, to save computational time and to check the FANOVA performance in the "worst case scenario" because the performance of an asymptotic test improves as sample size increases. The empirical type I error rates for the three association tests (FANOVA, FLM, and SKAT) were calculated based on 10,000 simulations. To further speed up computational time, FLM permutation-based p-values were still approximated based on $I = 999$ permutations because it is the smallest value of $I$ for which the empirical type I error rate estimate is unbiased for $\alpha = 0.001$. This value of $I$ can be calculated to provide equivalence of the left hand side to the desired nominal $\alpha$ level in

$$\Pr(\hat{p}_j \le \alpha) = \frac{\lfloor \alpha I \rfloor + 1}{I + 1}$$ [36], where $\hat{p}_j$ is the simulated p-value of the $j$th iteration and $\lfloor \; \rfloor$ is the "greatest integer contained in" or the "floor" function. The results of these additional simulations are presented in Table 2. As expected, the validity of all association tests holds.

We started our simulations by examining the impact of the proposed genotype relabeling method on power of functional procedures. The results are summarized in Figure (5). In Figure (4a), power was evaluated for $N = 50$ subjects based on a permutation test. In Figure (4b), power was evaluated for the same number of subjects but based on the parametric test. In Figure (5), thick lines represent power whenever genotype labeling was based on the count of minor alleles and slim lines represent power whenever genotype was relabeled to minimize the number of flips.

As expected, genotype relabeling drastically increases power of functional procedures if penalized cubic B-splines are used to obtain smooth curves (note the difference between thick and thin blue lines). In the case of a small number of cubic spline basis functions, the relabeling does not have much of an effect because the resulting smoothed function are "too smooth" and are insensitive to SNP relabeling. Similarly, in the case of a large number of cubic spline basis functions, the relabeling does not have much of an effect due to high oscillation in the fitted curves. Based on the comparison of Figure (4a) to Figure (4b), it is evident that statistical power of the parametric test is very similar to that of the permutation test. The power of FANOVA with the large number of basis functions is higher under the parametric test than for the permutation test, but it comes at the expense of an elevated type I error rate.

Taking these results into consideration, in the remainder of the article we used penalized cubic B-splines after genotype relabeling and the asymptotic test for the FANOVA method. We believe that penalized splines with the "optimal" amount of smoothing for each subject chosen through GCV is the most adequate way to represent genetic data out of the three different smoothing strategies that we considered. Additionally, parametric test has approximately the correct size and satisfactory power along with the best computational efficiency.

Simulations were then conducted to evaluate power of the three methods under the two different disease models (i.e., a model with both deleterious and protective effects and a model with deleterious effects) and varied percentage of causal variants. Figures (6) and (7) show power for each association testing procedure and $N = 50$. Under both disease models, statistical power of all procedures increases with the increase in the percentage of causal variants. The FANOVA test has the highest power for a small effect size ($\sigma_\beta = 0.25$) but for a larger effect size (i.e., $\sigma_\beta = 1$), FLM gains power and its performance is comparable or even higher than that of FANOVA. Both functional procedures have substantially higher power to detect an association than SKAT for $N = 50$. Figures (8) and (9) show power for $N = 500$ subjects and Figures (10) and (11) for $N = 1000$. Similar to the case of $N = 50$, FANOVA has the highest power when the magnitude of the effect size is small. SKAT gains power as the number of subjects increases and its performance becomes comparable to that of the functional methods.

To validate that FANOVA is also appropriate for GWAS data, we ran additional simulations in which we tested for an association a set of only common variants and a dichotomous phenotype. These results are available in Figures S1-S6 and show the same general pattern as for all variants. The only difference is that inclusion of rare variants increases statistical power of all of the procedures.

## Application

We used FANOVA, FLM and SKAT to test for an association between body mass index (BMI) and *ANGPTL3*, *ANGPTL4*, *ANGPTL5*, and *ANGPTL6*. The four genes are members of the *ANGPTL* family, with variants found contributing to plasma triglyceride levels [37]. Some recent studies suggest that *ANGPTL4* could be potentially associated with BMI [15,38]. To test for an association, we conducted a gene-based analysis by applying the three methods to each of the genes.

Although one of the strengths of functional approaches is that missing data are efficiently handled, software implementation of FLM by Luo et al. [15] does not provide straightforward handling of missing data. Therefore, in order to make a fair comparison between the methods, we handled missing genetic information by imputation using Bayesian linear regression model implemented in mice R package [39]. We excluded individuals with over 30% of missing variants (88 individuals for *ANGPTL 3*; 94 individuals for

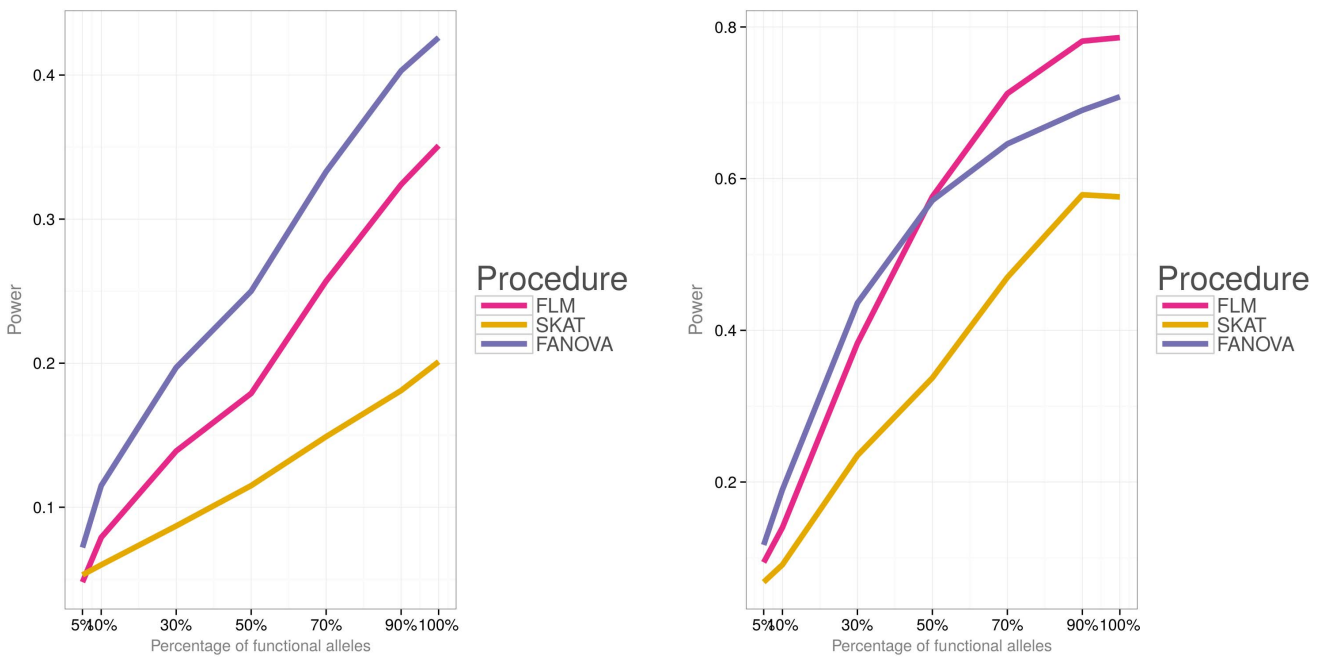

**Figure 10. Empirical power of the three methods for $N = 1000$ subjects, the first disease model (i.e., $\mu_\beta = 0$). Left panel:** small effect $\sigma_\beta = 0.05$. **Right panel:** large effect $\sigma_\beta = 0.1$
doi:10.1371/journal.pone.0105074.g010

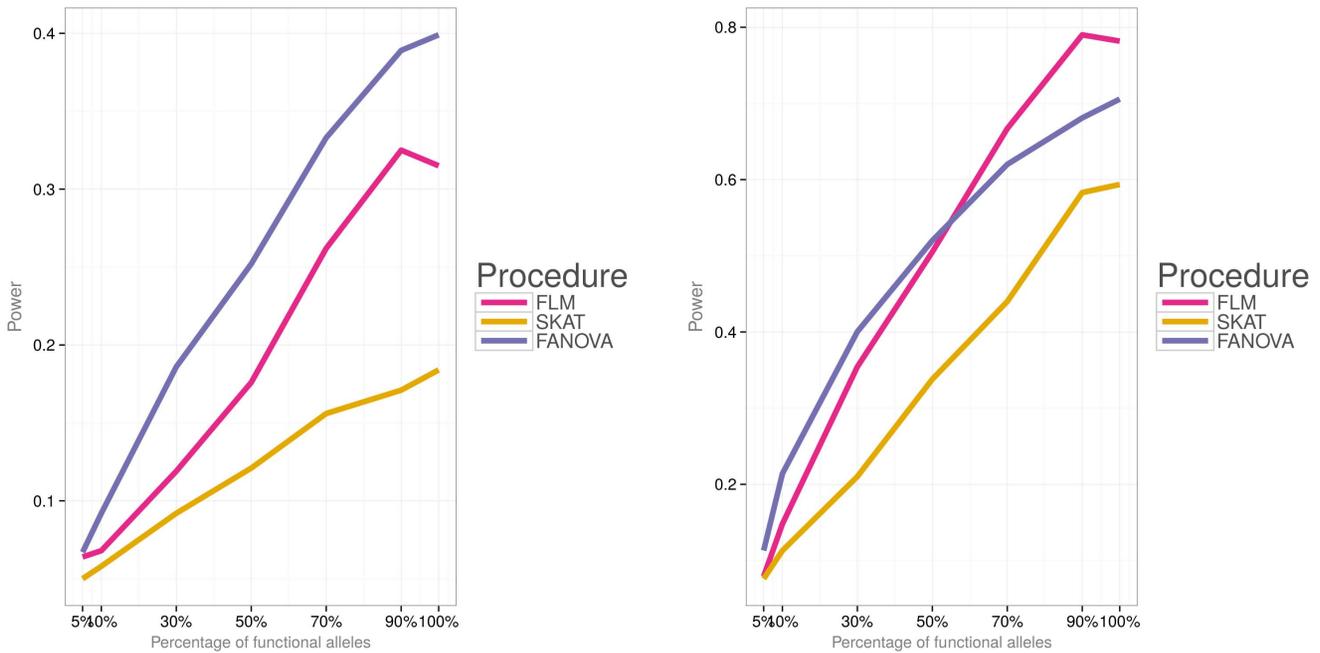

**Figure 11. Empirical power of the three methods for $N = 1000$ subjects, the second disease model (i.e., $\mu_\beta = 0.001$). Left panel:** small effect $\sigma_\beta = 0.05$. **Right panel:** large effect $\sigma_\beta = 0.1$.
doi:10.1371/journal.pone.0105074.g011

*ANGPTL 4*; 147 individuals for *ANGPTL 5*; and 103 individual for *ANGPTL 6*), as well as genetic variants that had information for less than 70% of subjects (none for *ANGPTL 3*; one for *ANGPTL 4*; 11 for *ANGPTL 5*; and none for *ANGPTL 6*). After multiple imputation and omitting variants with no variation, the total numbers of variants were 50, 70, 52, and 45 for *ANGPTL3*, *ANGPTL4*, *ANGPTL5*, and *ANGPTL6* genes, respectively. We examined the phenotypic association with BMI obesity by splitting the individuals into two groups: BMI≤30 and BMI>30. The sample sizes were: 1,307 subjects (*ANGPTL 3*), 1,596 subjects (*ANGPTL 4*), 1,739 subjects (*ANGPTL 5*), and 1,365 subjects (*ANGPTL 6*). Genomic positions were available for all variants but for those from the *ANGPTL 5* gene. For this gene, we assumed that sequencing variants were equally spaced over the genomic region. Table 3 summarizes results of our analysis. Both SKAT (p-value = 0.003) and FANOVA (p-value = 0.002) detected an association in *ANGPTL 4* with BMI obesity. In addition, FLM (p-value = 0.046) and FANOVA (p-value = 0.037) detected a marginal association between *ANGPTL 3* and BMI obesity. No association evidence has been found for *ANGPTL5* and *ANGPTL6*, which may suggest that these two genes make no contribution to BMI obesity, although the FANOVA method produced the smallest p-value of the approaches considered. After the Bonferroni adjustment for four tested genes, only the *ANGPTL4* association (by SKAT and FANOVA) remains significant. Nevertheless, the unadjusted p-values for *ANGPTL3* and *ANGPTL4* genes appear to confirm previous findings, considering that both genes have been linked to obesity and related phenotypes in prior studies that employed independent data sets [38,40–44].

## Discussion

We have presented a new method based on functional analysis of variance for detecting an association of multiple genetic variants with a dichotomous phenotypic trait. The FANOVA framework is a natural strategy for the case-control study design that exploits the correlation structure among variants. Functional data analysis (FDA) methods, including both FANOVA and FLM, effectively utilize position and LD information over sequenced data which can result in increased power over competing methods like SKAT. Furthermore, we have found that if the sample size or the magnitude of the majority of the effects is small, our procedure has substantially higher power than that of both FLM and SKAT. FANOVA surpasses these methods in a common situation when there are multiple susceptibility variants within a gene but statistical power to detect any particular variant by itself is insufficient.

We investigated different strategies for fitting functional objects to sequenced data. The typical FDA paradigm is that the smoother the original data are, the better, so that passing a smoother on each curve can effectively recover the true sample curves. If data jitter, the observations may contain additional random errors that need to be accounted for. As might be imagined, the typical FDA paradigm does not readily fit typical sequenced data. By using our flipping method of a reference allele, we tailored the genetic data to the FDA techniques which can result in a gain in power. To achieve better performance with our flipping method, the smoothing technique should be taken into careful consideration. With a typical approach based on a B-spline basis, a higher number of basis functions may result in an abundance of noise inherited by the smoother and subsequently may not lead to a power gain. To help overcome this noise, the number of basis functions can be truncated but the resulting fit may be overly smooth which would also not result in a gain in power. We recommend the use of penalized B-splines, in which the "optimal" amount of smoothing is based on the generalized cross validation algorithm. Our empirical findings demonstrated that our flipping method with the penalized B-splines greatly increased power of our procedure.

The functional analysis of variance method provides access to other potentially useful extensions. It allows for a follow-up test [45], in that once a genetic region associated with a phenotype is identified, it can be split into mutually-exclusive subregions to help focus on a subset of potentially functional variants and facilitate further identification of disease-causing variants. FANOVA also allows for a pair-wise comparison of phenotype levels [46], with which a researcher is able to contrast genotypes between pairs of a multilevel phenotype, either overall or within subregions where differences are identified. For example, in the majority of heritability studies of cognitive dysfunction, scaling of the observed trait is necessarily ordinal (e.g., from cognitively intact to cognitive dysfunction) [47] and it may be of interest to compare genetic differences between different levels of cognitive impairment. Future applications of the FANOVA methodology in this area would be able to borrow from these already developed tools.

A useful area for improvement would be an extension of our methodology to a multiple phenotype case as well as the ability to adjust for other covariates. If the additional phenotype and covariates are factors (e.g., sex), then multifactorial ANOVA for functional data provides a solution. Similar to Eq. 2, test statistics for the main-effect and for the interaction effect functions could be based on the $L^2$-norm of the functional sums of squares [25]. If the additional phenotype and covariates are continuous (e.g., age), then there are two possible solutions. First, if the purpose of the analysis is to investigate an association between a factor phenotype and a set of genetic variants, given any effects of continuous covariates, these additional effects can be treated as nuisance parameters and their effects can be eliminated prior to investigating the association of interest. A naive approach to eliminating these effects would be to regress a genotype function on continuous covariates and then to treat the resulting residual function as a dependent variable in regression on a factor phenotype. The problem with this approach is twofold: if the effect of a factor phenotype is non-null, the estimated effects of continuous covariates in the initial regression will be biased and power can be compromised; if the effect of a factor phenotype is null, but the factor phenotype and continuous covariates are collinear, the type I error rate can be conservative [48]. A possible solution, suggested by ter Braak [49] and Kennedy and Cade [48], is to regress both a factor phenotype and a genotype function on the continuous covariates. Then, our FANOVA methodology can be applied to the residualized factor alone with residualized genotype function. Second, if all of the effects are of interest (i.e., no nuisance parameters) the functional dependent variable can be related to univariate covarieties using a functional linear model but a new test statistic may be required. A possible solution to this problem was recently outlined by Reimherr and Nicolae [50].

We believe that functional analysis of variance is a promising method for efficient locus-wide inference. As we saw in our application, FANOVA is capable of delivering powerful results with interpretable conclusions. We hope that this paper will promote research on development and implementation of FDA methods for genetic studies.

## Supporting Information

**Figure S1 Empirical power of the three methods, only common variants, $N = 50$ subjects, first disease model (i.e., $\mu_\beta = 0$). Left panel: $\sigma_\beta = 0.25$. Right panel: $\sigma_\beta = 1$.**
(TIFF)

**Figure S2 Empirical power of the three methods, only common variants, $N = 50$ subjects, second disease model (i.e., $\mu_\beta = 0.05$). Left panel: $\sigma_\beta = 0.25$. Right panel: $\sigma_\beta = 1$.**
(TIFF)

**Figure S3 Empirical power of the three methods, only common variants, $N = 500$ subjects, first disease model (i.e., $\mu_\beta = 0$). Left panel: $\sigma_\beta = 0.05$. Right panel: $\sigma_\beta = 0.15$.**
(TIFF)

**Figure S4 Empirical power of the three methods, only common variants, $N = 500$ subjects, second disease model (i.e., $\mu_\beta = 0.01$). Left panel: $\sigma_\beta = 0.05$. Right panel: $\sigma_\beta = 0.15$.**
(TIFF)

**Figure S5 Empirical power of the three methods, only common variants, $N = 1000$ subjects, first disease model (i.e., $\mu_\beta = 0$). Left panel: $\sigma_\beta = 0.05$. Right panel: $\sigma_\beta = 0.1$.**
(TIFF)

**Figure S6 Empirical power of the three methods, only common variants, $N = 1000$ subjects, second disease model (i.e., $\mu_\beta = 0.001$). Left panel: $\sigma_\beta = 0.05$. Right panel: $\sigma_\beta = 0.1$.**
(TIFF)

## Acknowledgments

We thank Xiaowei Zhan, Dajiang Liu, and Jonathan Cohen for helping us access the Dallas Heart Study data set.

## Author Contributions

Analyzed the data: OAV. Contributed reagents/materials/analysis tools: OAV DVZ MCG CW QL. Wrote the paper: OAV QL.

## References

1. Raychaudhuri S (2011) Mapping rare and common causal alleles for complex human diseases. Cell 147: 57–69.
2. Hindorff LA, Sethupathy P, Junkins HA, Ramos EM, Mehta JP, et al. (2009) Potential etiologic and functional implications of genome-wide association loci for human diseases and traits. Proceedings of the National Academy of Sciences 106: 9362–9367.
3. Manolio TA, Collins FS, Cox NJ, Goldstein DB, Hindorff LA, et al. (2009) Finding the missing heritability of complex diseases. Nature 461: 747–753.
4. Ioannidis JP (2008) Why most discovered true associations are inflated. Epidemiology 19: 640–648.
5. Otto SP, Jones CD (2000) Detecting the undetected: Estimating the total number of loci underlying a quantitative trait. Genetics 156: 2093–2107.
6. Gibson G (2012) Rare and common variants: twenty arguments. Nature Reviews Genetics 13: 135–145.
7. Bodmer W, Bonilla C (2008) Common and rare variants in multifactorial susceptibility to common diseases. Nature genetics 40: 695–701.
8. Eichler EE, Flint J, Gibson G, Kong A, Leal SM, et al. (2010) Missing heritability and strategies for finding the underlying causes of complex disease. Nature Reviews Genetics 11: 446–450.
9. Goldstein DB, Allen A, Keebler J, Margulies EH, Petrou S, et al. (2013) Sequencing studies in human genetics: design and interpretation. Nature Reviews Genetics.
10. Madsen BE, Browning SR (2009) A groupwise association test for rare mutations using a weighted sum statistic. PLoS genetics 5: e1000384.
11. Wu M, Lee S, Cai T, Li Y, Boehnke M, et al. (2011) Rare-variant association testing for sequencing data with the sequence kernel association test. American Journal of Human Genetics 89: 82–93.
12. Neale BM, Rivas MA, Voight BF, Altshuler D, Devlin B, et al. (2011) Testing for an unusual distribution of rare variants. PLoS genetics 7: e1001322.
13. Lin DY, Tang ZZ (2011) A general framework for detecting disease associations with rare variants in sequencing studies. The American Journal of Human Genetics 89: 354–367.

14. Yashin AI, Wu D, Arbeev KG, Ukraintseva SV (2010) Joint influence of small-effect genetic variants on human longevity. Aging 2: 612–620.
15. Luo L, Zhu Y, Xiong M (2012) Quantitative trait locus analysis for next-generation sequencing with the functional linear models. J Med Genet 49: 513–524.
16. Ramsay J, Silverman B (2005) Functional Data Analysis. Springer, second edition.
17. Wood S (2006) Generalized Additive Models: An Introduction with R. Chapman & Hall/CRC Texts in Statistical Science.
18. Horvath L, Kokoszka P (2012) Inference for Functional Data with Applications. Springer.
19. Green P, Silverman BW (1994) Nonparametric Regression and Generalized Linear Models: A roughness penalty approach. Chapman & Hall/CRC Monographs on Statistics & Applied Probability.
20. Fan R, Wang Y, Mills JL, Wilson AF, Bailey-Wilson JE, et al. (2013) Functional linear models for association analysis of quantitative traits. Genetic Epidemiology 37: 726–742.
21. Hudson RR (1985) The sampling distribution of linkage disequilibrium under an infinite allele model without selection. Genetics 109: 611–631.
22. Kinnamon DD, Hershberger RE, Martin ER (2012) Reconsidering association testing methods using single-variant test statistics as alternatives to pooling tests for sequence data with rare variants. PLoS ONE 7: e30238.
23. Shen Q, Faraway J (2004) An F test for linear models with functional responses. Statistica Sinica 14: 1239–1257.
24. Cuevas A, Febrero M, Fraiman R (2004) An ANOVA test for functional data. Computational Statistics and Data Analysis 47: 111–122.
25. Zhang JT (2013) Analysis of Variance for Functional Data. Chapman & Hall/CRC Monographs on Statistics & Applied Probability.
26. D'Agostino RB (1972) Relation between the chi-squared and ANOVA tests for testing the equality of k independent dichotomous populations. The American Statistician 26: 30–32.
27. Lee S, Emond M, Bamshad M, Barnes K, Rieder M, et al. (2012) Optimal unified approach for rare variant association testing with application to small sample case-control whole-exome sequencing studies. Americal Journal of Human Genetics 91: 224–237.
28. Durbin RM, Altshuler DL, Durbin RM, Abecasis GAR, Bentley DR, et al. (2010) A map of human genome variation from population-scale sequencing. Nature 467: 1061–1073.
29. Ramsay JO, Wickham H, Graves S, Hooker G (2013) fda: Functional Data Analysis. URL http://CRAN.R-project.org/package=fda. R package version 2.3.6.
30. R Core Team (2013) R: A Language and Environment for Statistical Computing. R Foundation for Statistical Computing, Vienna, Austria. URL http://www.R-project.org/.
31. Wood S (2011) Fast stable restricted maximum likelihood and marginal likelihood estimation of semiparametric generalized linear models. Journal of the Royal Statistical Society 73: 3–36.
32. Lee S, Miropolsky L, Wu M (2013) SKAT: SNP-set (Sequence) Kernel Association Test. URL http://CRAN.R-project.org/package=SKAT. R package version 0.91.
33. Anderson MJ (2001) Permutation tests for univariate or multivariate analysis of variance and regression. Can J Fish Aquat 58: 626–639.
34. Reiss PT, Stevens HH, Shehzad Z, Petkova E, Milham MP (2010) On distance-based permutation tests for between-group comparisons. Biometrics 60: 636643.
35. Boos DD, Zhang J (2000) Monte carlo evaluation of resampling-based hypothesis tests. Journal of the American Statistical Association 95: 486–492.
36. Oden NL (1991) Allocation of effort in monte carlo simulation for power of permutation tests. Journal of the American Statistical Association 86: 1074–1076.
37. Romeo S, Pennacchio LA, Fu Y, Boerwinkle E, Tybjaerg-Hansen A, et al. (2007) Population-based resequencing of ANGPTL4 uncovers variations that reduce triglycerides and increase hdl. Nature genetics 39: 513–516.
38. Robciuc MR, Tahvanainen E, Jauhiainen M, Ehnholm C (2010) Quantitation of serum angiopoietin-like proteins 3 and 4 in a finnish population sample. Journal of lipid research 51: 824–831.
39. van Buuren S, Groothuis-Oudshoorn K (2011) mice: Multivariate imputation by chained equations in r. Journal of Statistical Software 45: 1–67.
40. Willer CJ, Sanna S, Jackson AU, Scuteri A, Bonnycastle LL, et al. (2008) Newly identified loci that influence lipid concentrations and risk of coronary artery disease. Nature genetics 40: 161–169.
41. Kathiresan S, Melander O, Guiducci C, Surti A, Burtt NP, et al. (2008) Six new loci associated with blood low-density lipoprotein cholesterol, high-density lipoprotein cholesterol or triglycerides in humans. Nature genetics 40: 189–197.
42. Kathiresan S, Willer CJ, Peloso GM, Demissie S, Musunuru K, et al. (2008) Common variants at 30 loci contribute to polygenic dyslipidemia. Nature genetics 41: 56–65.
43. Legry V, Bokor S, Cottel D, Beghin L, Catasta G, et al. (2009) Associations between common genetic polymorphisms in angiopoietin-like proteins 3 and 4 and lipid metabolism and adiposity in european adolescents and adults. The Journal of Clinical Endocrinology & Metabolism 94: 5070–5077.
44. Smart-Halajko MC, Kelley-Hedgepeth A, Montefusco MC, Cooper JA, Kopin A, et al. (2011) ANGPTL4 variants E40K and T266M are associated with lower fasting triglyceride levels in Non-Hispanic White Americans from the Look AHEAD Clinical Trial. BMC medical genetics 12: 89.
45. Vsevolozhskaya OA, Greenwood MC, Bellante GJ, Powell SL, Lawrence RL, et al. (2013) Combining functions and the closure principle for performing follow-up tests in functional analysis of variance. Computational Statistics & Data Analysis 67: 175–184.
46. Vsevolozhskaya OA, Greenwood MC, Holodov D (2014) Pairwise comparison of treatment levels in functional analysis of variance with application to erythrocyte hemolysis. The Annals of Applied Statistics: epub ahead of print.
47. Reynolds CA, Fiske A, Fratiglioni L, Pedersen NL, Gatz M (2006) Heritability of an age-dependent categorical phenotype: cognitive dysfunction. Twin Research and Human Genetics 9: 17–23.
48. Kennedy PE, Cade BS (1996) Randomization tests for multiple regression. Communications in Statistics – Simulations and Computation 25: 923–936.
49. ter Braak CJF (1992) Permutation versus bootstrap significance tests in multiple regression and ANOVA. Bootstrapping and related techniques 376: 79–85.
50. Reimherr M, Nicolae D (2014) A functional data analysis approach for genetic association studies. The Annals of Applied Statistic: epub ahead of print.